\documentclass[12pt,a4paper,oneside,notitlepage]{article}

\newif\ifpdf
\ifx\pdfoutput\undefined
   \pdffalse
\else
  \pdfoutput=1
  \pdftrue
\fi

\ifpdf
  \usepackage[pdftex]{graphicx}
  \usepackage[pdftex]{hyperref} 
\else
  \usepackage{graphicx}
  \usepackage{hyperref} 
\fi

\usepackage{c-series}
\usepackage{algorithmicx}
\usepackage{algorithm}
\usepackage{algpseudocode}

\usepackage{url}
\usepackage{amsfonts}

\usepackage{theorem}

\newtheorem{problem}{Problem}
\newtheorem{theorem}{Theorem}
\newtheorem{lemma}{Lemma}
\newenvironment{proof}{\emph{Proof. }}{\hfill $\Box$\vspace{2ex}

}

\begin{document}

%
%

\title{The Computational Complexity of Orientation Search Problems in Cryo-Electron Microscopy}
\author{Taneli Mielik\"ainen \and Janne Ravantti \and Esko Ukkonen}
\titlelinebreaks{The Computational Complexity \\ of Orientation Search Problems \\ in Cryo-Electron Microscopy}
\authorcontact{Taneli Mielik\"ainen \\
Department of Computer Science \\
University of Helsinki, Finland \\
\texttt{tmielika@cs.Helsinki.FI}
\and Janne Ravantti \\
Institute of Biotechnology and \\
Faculty of Biosciences \\
University of Helsinki, Finland \\
\texttt{ravantti@cs.Helsinki.FI}
\and Esko Ukkonen \\
Department of Computer Science \\
University of Helsinki, Finland \\
\texttt{ukkonen@cs.Helsinki.FI}
}

\reportyear{2004}
\reportno{3}
\reportmonth{June}
\CRClasses{F.2.2,I.4.5,J.3}
\printhouse{}
\reportpages{\pageref{contentsendpage} + \pageref{lastpage}}

\CRClassesLong{\newlength{\restline}
\setlength{\restline}{\textwidth}
\addtolength{\restline}{-13mm}
\parbox[t]{13mm}{F.2.2}
\vspace{1ex}
\parbox[t]{\restline}{Analysis of Algorithms and Problem Complexity: Nonnumerical Algorithms and Problems}
\parbox[t]{13mm}{I.4.5}
\parbox[t]{\restline}{Image Processing and Computer Vision: Reconstruction}
\parbox[t]{13mm}{J.3}
\parbox[t]{\restline}{Life and Medical Sciences: Biology and Genetics}
}               

\GeneralTerms{Algorithms, Theory}
\AdditionalKeyWords{Orientation Search, Line Arrangement, $NP$-hard, Inapproximable, Fixed-Parameter Intractable, Cryo-Electron Microscopy, Structural Biology}


%
%

\pagestyle{empty}

\makecover



\cleardoublepage
\abstractpagestart
In this report we study the problem of determining three-dimensional
orientations for noisy projections of randomly oriented identical
particles. The problem is of central importance in the tomographic
reconstruction of the density map of macromolecular complexes from
electron microscope images and it has been studied intensively for
more than 30 years.

We analyze the computational complexity of the orientation problem and
show that while several variants of the problem are $NP$-hard,
inapproximable and fixed-parameter intractable, some restrictions are
polynomial-time approximable within a constant factor or even solvable
in logarithmic space. The orientation search problem is formalized as
a constrained line arrangement problem that is of independent
interest. The negative complexity results give a partial justification
for the heuristic methods used in orientation search, and the positive
complexity results on the orientation search have some positive
implications also to the problem of finding functionally analogous
genes.

A preliminary version ``The Computational Complexity of Orientation
Search in Cryo-Electron Microscopy'' appeared in Proc.\ ICCS 2004,
LNCS 3036, pp.\ 231--238. Springer-Verlag 2004.
\abstractpageend

\label{contentsendpage}


%
%

\newpage
\pagenumbering{arabic}
\setcounter{page}{1}
\thispagestyle{plain}

\renewcommand{\sectionmark}[1]{%
\markright{\sectionname
\ \thechapter.\ #1}{}%
}

\pagestyle{headings}


\section{Introduction \label{s:i}}
Structural biology studies how biological systems are built.
Especially, determining three-dimensional electron density maps of
macromolecular complexes, such as proteins or viruses, is one of the
most important tasks in structural biology~\cite{f}.

Standard techniques to obtain three-dimensional density maps of such
particles (at atomic resolution) are by X-ray diffraction
(crystallography) and nuclear magnetic resonance (NMR) studies.
However, X-ray diffraction requires that the particles can form
three-dimensional crystals and the applicability of NMR is limited to
relatively small particles~\cite{csrsm}. For example, there are many
well-known viruses that do not seem to crystallize and are too large
for NMR techniques. (To the best of our knowledge NMR techniques can
be currently applied only up to size of $1$ MDa~\cite{fw} while
viruses are typically at least ten times larger.)

A more flexible way to reconstruct density maps is offered by
cryo-electron microscopy~\cite{cdk,f}. Currently the resolution of the
cryo-electron microscopy reconstruction is not quite as high as
resolutions obtainable by crystallography or NMR but it is improving
steadily.

\begin{figure} \centering
\includegraphics[width=\textwidth]{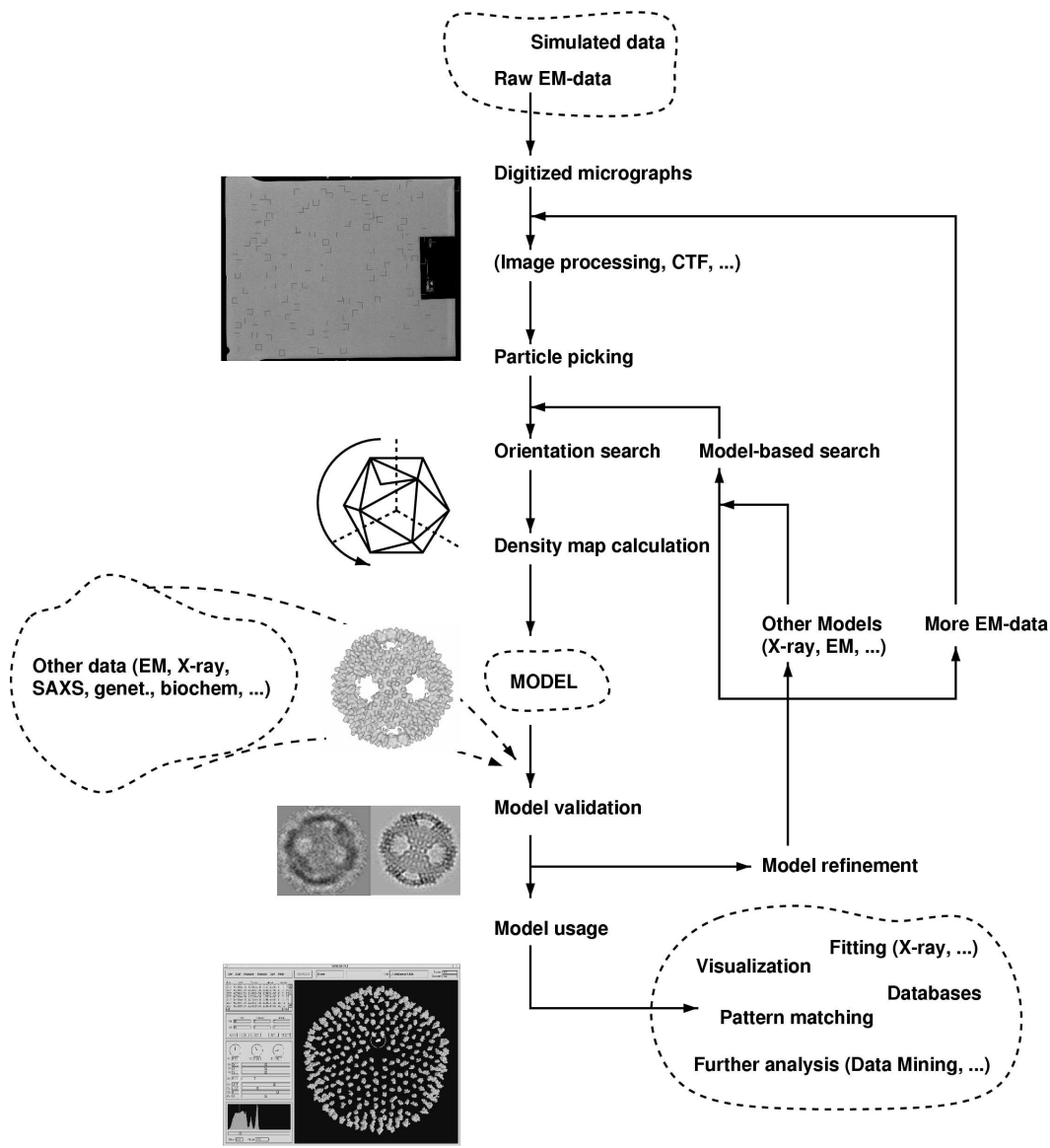}
\caption{The reconstruction process. \label{f:reconstruction}}
\end{figure}

Reconstruction of density maps by cryo-electron microscopy consists of
the following subtasks:
\begin{description}
\item[Specimen preparation.] A thin layer of water containing a large
number of identical particles of interest is rapidly plunged into
liquid ethane to freeze the specimen very quickly.  Quick cooling
prevents water from forming regular structures \cite{f}.  Moreover,
the particles get frozen in random orientations in the iced specimen.
\item[Electron microscopy.] The electron microscope produces an image
representing a two-dimensional projection of the iced specimen. This
image is called a \emph{micrograph}. Unfortunately the electron beam
of the microscope rapidly destroys the specimen so getting accurate
images from it is not possible.
\item[Particle picking.] Individual projections of particles are
extracted from the micrograph. There are efficient methods to do that,
see e.g.~\cite{krvub,ng}. The number of projections obtained may be
thousands or even more.
\item[Orientation search.] The orientations (i.e., the projection
directions for each extracted particle) for the projections are
determined.  There are a few heuristic approaches for finding the
orientations.  For further details, see Section~\ref{s:op}.
\item[Reconstruction.] If the orientations for the projections are
known then quite standard tomography techniques can be applied to
construct the three-dimensional electron density map from the
projections~\cite{f}.
\end{description}
For a more broader view to the reconstruction process, see
Figure~\ref{f:reconstruction}.

In this report we study the computational complexity of the
orientation search problem which is currently the major bottleneck in
the reconstruction process.  On one hand we show that several variants
of the task are computationally very difficult. This justifies (to
some extent) the heuristic approaches used in practice.  On the other
hand we give exact and approximate polynomial-time algorithms for some
special cases of the task that are applicable e.g.\ to the seemingly
different task of finding functionally analogous genes~\cite{hl}.

The rest of this report is organized as follows. In Section~\ref{s:op}
the orientation search problem is described.  Section~\ref{s:co}
analyzes the computational complexity and approximability of the
orientation search problem. As an abstract formulation of the search
problem we use certain constrained line arrangement problems that are
of independent interest. The report is concluded in Section~\ref{s:c}.

\section{The Orientation Search Problem \label{s:op}}
A \emph{density map} is a mapping $D : \mathbb{R}^3 \to \mathbb{R}$
with a compact support.  An \emph{orientation} $o$ is a rotation of
the three-dimensional space and it can be described e.g.\ by a
three-dimensional rotation matrix.

A \emph{projection} $p$ of a three-dimensional density map $D$ to
orientation $o$ is the integral
\begin{displaymath}
p\left(x,y\right)=\int_{-\infty}^{\infty} D\left(R_o\left[x,y,z\right]^T\right) dz
\end{displaymath}
where $R_o$ is a three-dimensional rotation matrix, i.e., the mass of
$D$ is projected on a plane passing through the origin and determined
by the orientation $o$.

Projections of physical densities can be produced e.g.\ by X-rays or
electron microscopy.  In practice, the density maps are usually
represented as three-dimensional regular grids of finite-precision
numbers depending on the accuracy of the scanning device but in this
report we do not need to consider the actual representations of
projections or density maps.

Based on the above definitions, the orientation search task is, given
projections $p_1,\ldots,p_n$ of the same underlying but unknown
density map $D$ to find good orientations $o_1,\ldots,o_n$ for them.
There are several heuristic definitions of what are the good
orientations for the projections.

One possibility is to choose those orientations that determine a good
density map although it might not be obvious what a good density map
is nor how it should be constructed from oriented projections.  A
standard solution is to compare how well the given projections fit to
the projections of the reconstructed density map.  This kind of
definition of good orientations suggests an Expectation
Maximization-type procedure of repeatedly finding the best model for
fixed orientations and the best orientations for a fixed model, see
e.g.~\cite{bc-m,dj,jmcb,ljtcsm,yzd}.  Due to the strong dependency on
the reconstruction method, it is not easy to say analytically much
(even whether it converges) about this approach in general.  In
practice, this approach to orientation search works successfully if
there is an approximate density map of the particle available to be
used as an initial model.

\begin{figure}
\includegraphics[width=\textwidth]{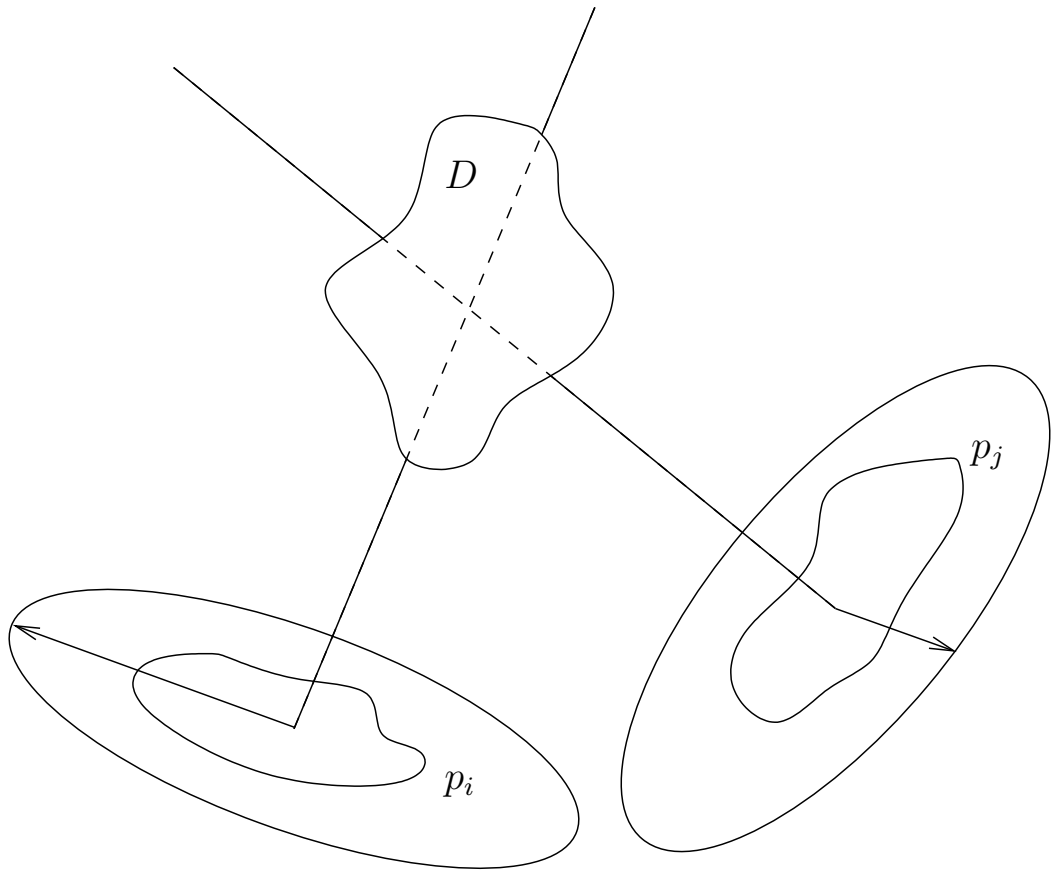}
\caption{Two projections of density $D$. \label{f:projections}}
\end{figure}

\begin{figure}
\includegraphics[width=\textwidth]{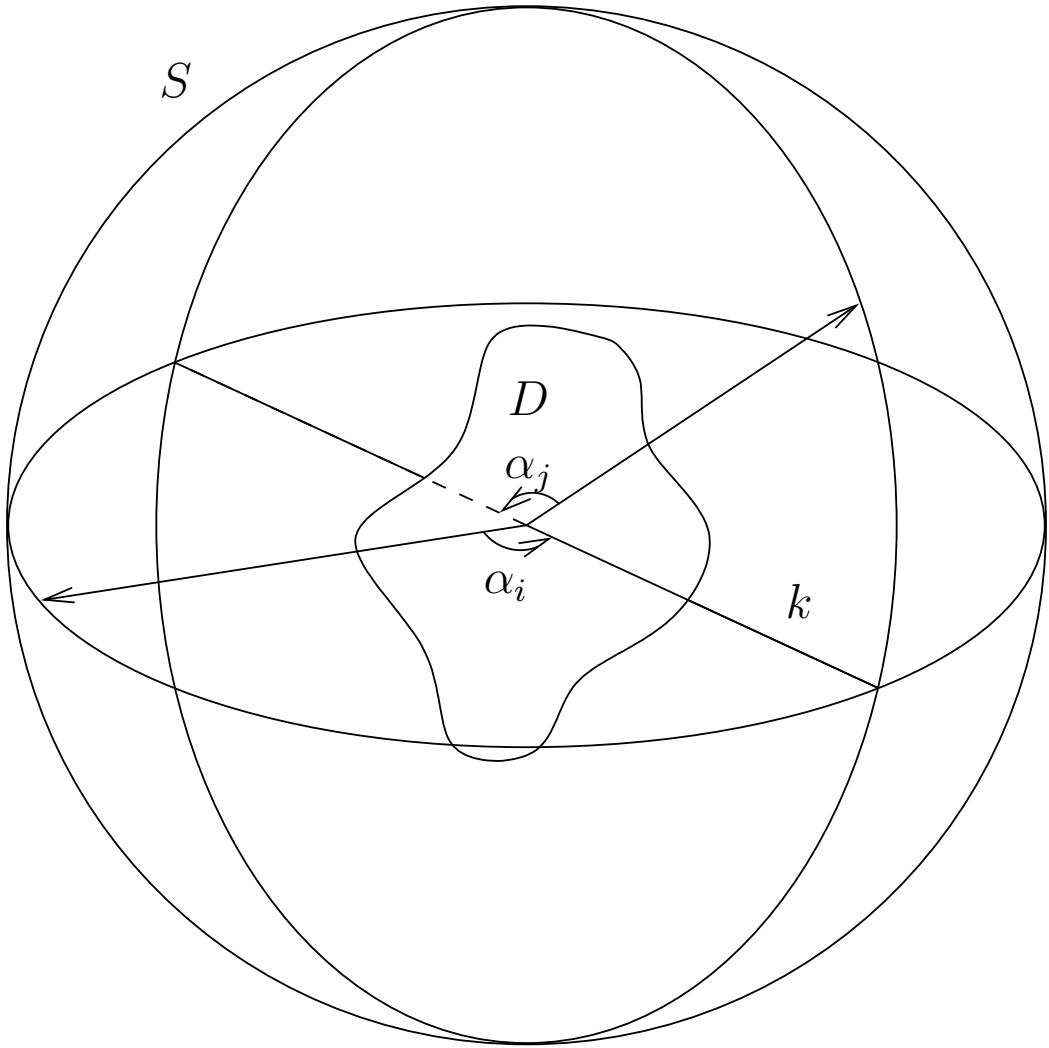}
\caption{Two projection directions presented as great circles and their common line $k$ specified with the rotation angles $\alpha_i$ and $\alpha_j$ in the internal coordinate systems of the two circles. \label{f:directions}}
\end{figure}

The orientations can be determined also by \emph{common
lines}~\cite{bof}: Let $p_i$ and $p_j$ be projections of a density map
$D$ onto planes corresponding to orientations $o_i$ and $o_j$,
respectively; see Figure~\ref{f:projections}.  All one-dimensional
projections of $D$ onto a line passing through the origin in the plane
corresponding to the orientation $o_i$ ($o_j$) can be computed from
the projection $p_i$ ($p_j$); this collection of projections of $p_i$
($p_j$) is also called the \emph{sinogram} of $p_i$ ($p_j$).  As the
two planes intersect, there is a line for which the projections of
$p_i$ and $p_j$ agree.  This line (which actually is a vector since
the one dimensional projections are oriented, too) is called the
\emph{common line} of $p_i$ and $p_j$; Figure~\ref{f:directions}.

If the projections are noiseless then already the pairwise common
lines of three projections determine the relative orientations of the
projections in three-dimensional space uniquely (except for the
handedness) provided that the possible symmetries of the particle are
taken into account.  Furthermore, this can be computed by only few
arithmetic and trigonometric operations~\cite{h-ar}.

However, the projections produced by the electron microscope are
extremely noisy and so it is highly unlikely that two projections have
one-dimensional projections that are equal.  In this case it would be
natural to try to find the best possible approximate common lines,
i.e., a pair of approximately equal rows from the sinograms for the
two projections.  Several heuristics for the problem have been
proposed~\cite{bcl,bls,cbst,cdk,fbcb,l,pzf,tc,h-ar}.  However, they
usually assume that the density map under reconstruction is highly
symmetric which radically improves the signal-to-noise ratio.  In
Section~\ref{s:co} we partially justify the use of heuristics by
showing that many variants of the orientation search problem are
computationally very difficult.

\section{The Complexity of Orientation Search \label{s:co}}

In this section we show that finding good orientations using common
lines is computationally very difficult in general but it has some
efficiently solvable special cases.  The results are described in
three phases: First, we consider the decision versions of the
orientation search problem.  Second, we study the approximability of
several optimization variants.  Finally, we examine the parameterized
(in)tractability of the problem.

We would like to point out that some of the results are partially
similar to the results of Hallett and Lagergren~\cite{hl} for their
problem \textsc{Core-Clique} that models the problem of finding
functionally analogous genes.  However, our problem of finding good
orientations based on common lines differs from the problem of finding
functionally analogous genes, e.g., by its geometric nature and by its
very different application domain.  Furthermore, we provide relevant
positive results for finding functionally analogous genes: we describe
an approximation algorithm with guaranteed approximation ratio of
$2\beta\left(1-o(1)\right)$, if the distances between genes adhere to
the triangle inequality within a factor $\beta$.

\subsection{Decision Complexity \label{s:dc}}
As mentioned in Section~\ref{s:op}, the pairwise common lines cannot
be detected reliably when the projections are very noisy.  A natural
relaxation is to allow several common line candidates for each pair of
projections.  In this section we study the problem of deciding whether
there exist common lines in given sets of pairwise common lines that
determine consistent orientations.  We show that some formulations are
$NP$-complete in general but there are nontrivial special cases that
are solvable in nondeterministic logarithmic space.  (For further
information about computational complexity and complexity classes of
decision problems, see e.g.~\cite{p}.)

The common lines-based orientation search problem can be modeled at a
high level as the problem of finding an $n$-clique from an
$n,m$-partite graph $G=\left(V_1,\ldots,V_n,E\right)$, i.e., a graph
consisting independent sets $V_1,\ldots,V_n$ of size $m$.
\begin{problem}[$n$-clique in an $n,m$-partite graph] \label{p:ncnm}
Given an $n,m$-partite graph $G=\left(V_1,\ldots,V_n,E\right)$, decide
whether there is an $n$-clique in $G$.
\end{problem}

Problem~\ref{p:ncnm} can be interpreted as the orientation search
problem in the following way: each group $V_i$ describes the possible
orientations of the projection $p_i$ and each edge connecting two
oriented projections says that the projections in the corresponding
orientations are consistent with each other.

On one hand already three different orientations for each projection
can make the problem $NP$-complete:
\begin{theorem}
Problem~\ref{p:ncnm} is $NP$-complete if $m \geq 3$. \label{t:ncnm3}
\end{theorem}
\begin{proof}
Clearly, the problem is in $NP$ since one can check in polynomial time
in $|G|$ whether a given subset of the vertices of $G$ forms an
$n$-clique.

We show the $NP$-hardness of Problem~\ref{p:ncnm} by reduction from
the graph $k$-colorability problem:
\begin{problem}[graph $k$-colorability~\cite{p}] \label{p:gc}
Given a graph $G=\left(V,E\right)$ and a positive integer $k$, decide
whether $G$ is $k$-colorable, i.e., whether there is a mapping $f : V
\to \left\{ 1,\ldots,k \right\}$ such that if $\left\{ u,v \right\}
\in E$ then $f\left(u\right)\neq f\left(v\right)$.
\end{problem}

Let $G'=(V',E')$ be the graph that we would like to color with $k$
colors.  The polynomial-time reduction to a corresponding instance
$G=(V_1,\ldots,V_n,E)$ of Problem~\ref{p:ncnm} is as follows. For each
vertex $i \in V'$ there is a group $V_i$ consisting of $k$ vertices
$v^1_i,\ldots,v^k_i$.  Each vertex in $V_i$ corresponds to one
coloring of the vertex $i \in V'$.  There is an edge $\{ v_i,v_j \}
\in E, v_i \in V_i, v_j \in V_j,i\neq j,$ if and only if $\{i,j\}
\notin E'$ or $v_i$ and $v_j$ are of different color.

Clearly, the graph $G'=\left(V',E'\right)$ is $k$-colorable if and
only if there is an $n$-clique in the corresponding $n,k$-partite
graph $G=\left(V_1,\ldots,V_n,E\right)$.  The members of groups $V_i$
that correspond to a coloring form an $n$-clique in $G$.
\end{proof}

On the other hand the problem can be solved in nondeterministic
logarithmic space if the number of orientations for each projection is
at most two:
\begin{theorem}
Problem~\ref{p:ncnm} is $NL$-complete if $m \leq 2$. \label{t:ncnm2}
\end{theorem}\begin{proof}
The problem is in $NL$ since it can be reduced in logarithmic space to
the $m$-satisfiability problem with $m \leq 2$ that is an
$NL$-complete problem:
\begin{problem}[$m$-satisfiability \cite{p}] \label{p:msat}
Given a set $U$ of boolean variables and a set $C$ of clauses $c \in
C,|c|\leq m$, decide whether there is a truth value assignment $f : U
\to \{ 0,1 \}$ that satisfies all clauses in $C$, i.e., whether there
is a truth value assignment $f$ that sets at least one
literal\footnote{Recall that literals are just boolean formulas of
type $x=0$ and $x=1$ where $x$ is a variable.}  true in each clause of
$C$.
\end{problem}

Note first that any instance of the problem with $m \leq 2$ can be
trivially reduced to the case with $m=2$.  The reduction from
Problem~\ref{p:ncnm} with $m = 2$ to Problem~\ref{p:msat} with $m=2$
is as follows. Let the instance of Problem~\ref{p:ncnm} be
$G=\left(V_1,\ldots,V_n,E\right)$ and the instance of
Problem~\ref{p:msat} $\left(U,C\right)$.  For each group
$V_i=\{v^0_i,v^1_i\}$ there is a boolean variable $u_i$ whose truth
value assignments $u_i=0$ and $u_i=1$ correspond to vertices $v^0_i$
and $v^1_i$, respectively.  The set $C$ contains a clause $u_i=(1-a)^2
\lor u_j=(1-b)^2$ if and only if $\{ v^a_i , v^b_j \} \notin E$.

If there is a truth assignment $f$ satisfying all clauses in $C$ then
the vertices corresponding to the truth value assignments form an
$n$-clique $V'$ in $G$: Assume contrary that the truth value
assignment $f$ satisfies all clauses in $C$ but the corresponding set
$V'$ of $n$ vertices does not form an $n$-clique.  Then there are at
least two vertices $v_i^a$ and $v_j^b$ in $V'$ such that $\{
v_i^a,v_j^b\} \notin E$.  But then $C$ contains a clause $u_i=(1-a)^2
\lor u_j=(1-b)^2$ which the truth value assignment $f$ does not
satisfy.  If no truth value assignment $f$ satisfies all clauses in
$C$ then in any set $V'$ of $n$ vertices there are at least two
vertices $v_i^a$ and $v_j^b$ such that $\{ v_i^a,v_j^b\} \notin E$.

Thus, the graph $G$ contains an $n$-clique if and only if there if a
truth value assignment $f$ that satisfies all clauses in $C$.

The problem is also $NL$-hard since Problem~\ref{p:msat} with $m=2$
can be reduced to it in logarithmic time in a similar way.
\end{proof}

The formulation of the orientation search problem as
Problem~\ref{p:ncnm} seems to miss some of the geometric nature of the
problem.  As a first step toward the final formulation, let us
consider the problem of finding a constrained line arrangement, the
constraint being that any two lines of the arrangement are allowed to
intersect only at a given set of points, each such set being of size
$\leq l$:
\begin{problem}[$l$-constrained line arrangement] \label{p:cla}
Given sets $P_{ij} \subset \mathbb{R}^2,|P_{ij}|\leq l,1 \leq i < j
\leq n$, decide whether there exist lines $L_1,\ldots,L_n$ in
$\mathbb{R}^2$ such that $L_i$ and $L_j$ intersect only at some $p \in
P_{ij}$ for all $1 \leq i < j \leq n$.
\end{problem}

This problem has some interest of its own since line arrangements are
one of the central concepts in computational and discrete
geometry~\cite{e-acg,m-ldg}. If we require that the lines are in
general position, i.e., that they are not parallel nor they intersect
in same points, then we get the following hardness result:
\begin{theorem}
Problem~\ref{p:cla} is $NP$-complete if $l \geq 9$. \label{t:cla9}
\end{theorem}\begin{proof}
The problem is in $NP$ for all $l \geq 0$ since it can be checked in
polynomial time whether there are lines $L_1,\ldots,L_n$ such that
$L_i$ and $L_j$ intersect at $p_{ij}$ for each $1 \leq i < j \leq n$.

The $NP$-hardness of the problem can be shown by a polynomial-time
reduction from Problem~\ref{p:ncnm} as follows.  Let
$G=\left(V_1,\ldots,V_n,E\right)$ be the instance of
Problem~\ref{p:ncnm}.  For each vertex $v_{i,a} \in V_i$ we have a
line $L_{i,a}$.  Set $P_{ij}$ contains the intersection point of lines
$L_{i,a}$ and $L_{j,b}$ if and only if $\{v_{i,a},v_{j,b}\} \in E$.
We can use this reduction if we are able to find $nm$ lines on plane
in \emph{general position} (for discussion on what being in general
position means, see \cite{m-ldg}).  Actually, it is sufficient to
require that
\begin{enumerate}
\item
no two lines are parallel,
\item
no three lines intersect in the same point, and
\item
if $p_{ij_1} \in P_{ij_1}$, $p_{ij_2} \in P_{ij_2}$ and $p_{ij_3} \in
P_{ij_3}$ are on same line then this line is one of the lines
$L_{i,a}$.
\end{enumerate}

Non-vertical lines $y=gx+h$ can be mapped to points $(g,h) \in
\mathbb{R}^2$ and vice versa.  The $nm$ lines can be generated by
considering the pairs $(g,h) \in \mathbb{N}^2$ of positive integers in
lexicographical order $\prec$: $(g_1,h_1) \prec (g_2,h_2)$ if and only
if $g_1<g_2 \lor \left( g_1=g_2 \land h_1<h_2\right)$; and choosing
some points $(g,h)$ according to rules that are equivalent to the
above rules for lines.  The rules for choosing the points are:
\begin{enumerate}
\item
each chosen point has a unique first coordinate $g$; we call $g$ the
\emph{column index} of the point,
\item
no line passes through three chosen points, and
\item
three lines, each passing through two chosen points, can intersect in
the same point only if that point is chosen, too.
\end{enumerate}

We still have to show that it is sufficient to consider only a
polynomial number of points in $\mathbb{N}^2$ in order to find $nm$
points that satisfy the given requirements.  Let the number of chosen
points at certain stage of the construction to be $k$ with one point
chosen from each column $0, \ldots, k-1$.  Then the maximum number of
the points we have to consider at column $k$ before finding a feasible
point can be bounded above polynomially in $n$ and $m$ as follows:
\begin{itemize}
\item
Exactly $\left(k \atop 2\right)$ lines can be drawn passing through at
least two chosen points.  These lines make at most $\left(k \atop 2
\right)$ points in the column $k$ infeasible.
\item
Two points on a plane span a line uniquely.  Any four chosen points
span two different lines $L'_i$ and $L'_j$ and there are exactly
$\left(k \atop 4 \right)$ such pairs of lines.  Each of the other
$k-4$ chosen points can span at most one line with the points in the
column $k$ that passes through the intersection point of the lines
$L'_i$ and $L'_j$.  Thus, the number of points in the column $k$ that
are infeasible due to this is at most $(k-4)\left(k \atop 4 \right)$.
\end{itemize}

Thus, the number of points in column $k$ that have to be considered
before finding the first point that does not violate our selection
rules and hence can be chosen as the $k+1$:st point is at most
\begin{displaymath}
\left(k \atop 2 \right)+\left(k-4\right)\left(k \atop 4\right)
\end{displaymath}
which is clearly polynomial in $nm$ when $k \leq nm$.
\end{proof}

The result can be slightly improved if we relax the general position
requirement used in Theorem~\ref{t:cla9}, e.g., if we allow also
parallel lines in the arrangement:
\begin{theorem}
Problem~\ref{p:cla} is $NP$-complete if $l \geq 6$. \label{t:cla6}
\end{theorem}
\begin{proof}
The problem is in $NP$ as noted in the proof of Theorem~\ref{t:cla9}.

The $NP$-hardness of the problem can be shown by reduction from
Problem~\ref{p:msat} as follows. Given an instance $\left(C,U\right)$
of the $m$-satisfiability problem, we construct point sets $P_{ij}$
for $1 \leq i \leq \left|U\right|$ and $1 \leq j \leq
\left|C\right|$. This is done by representing the variables and
clauses by suitable line arrangements and constraining their
intersection points. Each boolean variable $u_i \in U$ is represented
by two vertical lines $L^0_i$ and $L^1_i$ representing the truth value
assignments $u_i=0$ and $u_i=1$, respectively.  Each clause $c_j \in
C$ is represented by $|c_j|$ horizontal lines $L_{j,1},
\ldots,L_{j,|c_j|}$. The intersection point of lines $L_{i,a}$ and
$L_{j,b}$ corresponding to the truth value assignment $u_i=a$ and the
$b$th literal in the clause $c_j$ is in $P_{ij}$ if and only if the
truth value assignment $u_i=a$ does not falsify the $b$th literal in
$c_j$ which fixes sets $P_{ij}$. These lines are placed on plane in
such way that all vertical lines have different horizontal coordinates
and all horizontal lines have different vertical coordinates.

Without loss of generality, we assume that $\left|U\right| > m$ and
$\left|C\right| > 2$. This ensures that all lines that are spanned by
the points in sets $P_{ij}$ and correspond to the clauses must be
horizontal and all lines that correspond to the variables must be
vertical in any feasible line arrangement corresponding to a
satisfying truth assignment.

If there is a satisfying truth assignment $f$ for the set $C$ of
clauses then the lines of the corresponding line arrangement intersect
in the allowed points that belong to the sets $P_{ij}$. If there are
lines intersecting only at the allowed points then the vertical lines
uniquely determine a truth value assignment $f$ that satisfies all
clauses in $C$.

Thus, the lines can be arranged on plane in such way that they
intersect only at allowed intersection points in sets $P_{ij}$ if and
only if the there is a truth value assignment satisfying all clauses
in $C$.  Furthermore, if the size of the largest clause is $m$ then
the size of the largest set $P_{ij}$ is at most $2m=l$.  As
Problem~\ref{p:msat} is $NP$-complete when $m \geq 3$,
Problem~\ref{p:cla} is $NP$-complete when $l\geq 6$.
\end{proof}

However, the orientation search is not about arranging lines on the
plane but great circles on the (unit) sphere $S=\left\{
\left(x,y,z\right) \in \mathbb{R}^3 : x^2+y^2+z^2=1 \right\}$ as the
orientations and the great circles are obviously in one-to-one
correspondence.  Thus, we should study the great circle arrangements:
\begin{problem}[$l$-constrained great circle arrangement] \label{p:cgca}
Given sets $P_{ij} \subset S_+=\left\{ \left(x,y,z\right) \in S : z
\geq 0 \right\}, \left|P_{ij}\right| \leq l, 1 \leq i < j \leq n$,
decide whether there exist great circles $C_1,\ldots,C_n$ on $S$ such
that $C_i$ and $C_j$ intersect on $S_+$ only at some $p \in P_{ij}$
for all $1 \leq i < j \leq n$.
\end{problem}

It can be shown that the line arrangements and great circle
arrangements are equivalent through the stereographic projection
\cite{e-acg}:
\begin{theorem}
Problem~\ref{p:cgca} is as difficult as
Problem~\ref{p:cla}. \label{t:cgca}
\end{theorem}
\begin{proof}
Great circles on a sphere can be mapped to lines on a plane by the
central projection and lines on a plane to great circles on a sphere
by its inverse~\cite{by}.
\end{proof}

Still, our problem formulation is lacking some of the important
ingredients of the orientation search problem: it is not possible to
express at this stage of the orientation search the common line
candidates by giving the allowed pairwise intersection points on the
sphere $S$, i.e., in some globally fixed coordinate system.  Rather,
one can represent a common line only in the internal coordinates of
the two great circles that correspond to the two projections
intersecting.  Each coordinate is in fact an angle giving the rotation
angle of the common line on the projection as depicted in
Figure~\ref{f:directions}.  Hence the representation is a pair of
angles:

\begin{problem}[locally $l$-constrained great circle arrangement on sphere]\label{p:lcgca}
Given sets $P_{ij} \subset [0,2\pi) \times[0,2\pi),|P_{ij}| \leq l, 1
\leq i < j \leq n$, decide whether there exist great circles
$C_1,\ldots,C_n$ on $S$ such that $C_i$ and $C_j$ intersect only at
some $p \in P_{ij}$ for all $1 \leq i < j \leq n$, where $p$ defines
the angles of the common line on $C_i$ and $C_j$.
\end{problem}

Also this problem can be shown to be equally difficult to decide:
\begin{theorem}
Problem~\ref{p:lcgca} is $NP$-complete if $l \geq 6$. \label{t:lcgca}
\end{theorem}
\begin{proof}
The problem is in $NP$ since it is possible to check in polynomial
time in the total number of possible local intersection points whether
a given set of local intersection points is realizable.

The $NP$-hardness of the problem can be obtained from the proofs of
Theorem~\ref{t:cla6} and Theorem~\ref{t:cgca}. Indeed, all great
circles corresponding to the horizontal lines in the corresponding
line arrangement are forced to be parallel by their common
intersection point. Similarly, all great circles corresponding to the
vertical lines are forced to be parallel by their common intersection
point.
\end{proof}

Thus, deciding whether there exist consistent orientations seems to be
difficult in general.

\subsection{Approximability \label{s:a}}
As finding a consistent orientation for the projections is by the
results of Section~\ref{s:dc} difficult, we should consider also
orientations that may determine orientations only for a large subset
of the projections or resort to common lines that are as good as
possible.

A simple approach to consider consistent orientations for large
subsets of the projections is to look for large cliques in the
$n,m$-partite graph $G=\left(V_1,\ldots,V_n,E\right)$ instead of
exactly $n$-cliques.  In the world of orientations this means that
instead of finding consistent orientations for all projections we look
for consistent orientations for as many projections as we are able to
and neglect the other projections.

Containing a clique is just one example of a property a graph can
have. Also other graph properties might be useful. Thus we can
formulate the problem in a rather general form as follows:
\begin{problem}[Maximum subgraph with property $P$ in an $n,m$-partite graph] \label{p:msn}
Given an $n,m$-partite graph $G=\left(V_1,\ldots,V_n,E\right)$, find
the largest $V' \subseteq V_1 \cup \ldots \cup V_n$ such that the
induced subgraph satisfies the property $P$ and $\left|V' \cap
V_i\right| \leq 1$ for all $1 \leq i \leq n$.
\end{problem}

This resembles the following fundamental graph problem in
combinatorial optimization and approximation algorithms:
\begin{problem}[Maximum subgraph with property $P$ \cite{acgkmp}] \label{p:ms}
Given a graph $G=\left(V,E\right)$, find the largest $V' \subseteq V$
such that the induced subgraph satisfies the property $P$.
\end{problem}

It is not very difficult to see that the two problems are equivalent:
\begin{theorem}
Problem~\ref{p:msn} is as difficult as
Problem~\ref{p:ms}. \label{t:msnms}
\end{theorem}
\begin{proof}
On the one hand, Problem~\ref{p:msn} is a special case of
Problem~\ref{p:ms} with a restricted graph structure and with the
additional condition $\left|V' \cap V_i\right|\leq1$ for all $i$ which
can be included in the property $P$.  On the other hand,
Problem~\ref{p:ms} is a special case of Problem~\ref{p:msn} with
singleton groups $V_1,\ldots,V_{|C|}$.
\end{proof}

Problem~\ref{p:ms} is very difficult w.r.t.\ several
properties~\cite{acgkmp}.  By Theorem~\ref{t:msnms}, these results
generalize to Problem~\ref{p:msn}.  Hence, for example, finding the
maximum clique from the $n,m$-partite graph cannot be approximated
within ratio $n^{1-\epsilon}$ for any fixed $\epsilon >0$~\cite{h-c}.
Note that the approximation ratio $n$ can be achieved trivially by
choosing any of the vertices in $G$ which is always a clique of size
$1$.

In practice the techniques for finding common lines or common line
candidates actually evaluate all potential common lines of two
projections (that is, all relative orientations of the two projections
with respect to each other are in effect considered) and give them a
score which typically is the distance between the two sinogram rows
corresponding to potential common line.  Thus, we could assume that
there is always at least one feasible solution and study the following
problem:
\begin{problem}[Minimum weight $n$-clique in a complete $n,m$-partite graph] \label{p:mwn}
Given a complete $n,m$-partite graph $G=\left(V_1,\ldots,V_n,E\right)$
and a weight function $w: E \to \mathbb{N}$, find $V' \subset V_1 \cup
\ldots V_n$ such that the weight $\sum_{u,v \in V', u \neq v}
w(\left\{u,v\right\})$ is minimized and $\left|V' \cap
V_i\right| \leq 1$ for all $1 \leq i \leq n$.
\end{problem}

Unfortunately, it turns out that in this case the situation is
extremely bad:
\begin{theorem}
Problem~\ref{p:mwn} with $m \geq 3$ is not polynomial-time
approximable within $2^{n^k}$ for any fixed $k>0$ if $P \neq
NP$. \label{t:mwn3}
\end{theorem}
\begin{proof}
If Problem~\ref{p:mwn} were approximable within $2^{n^k}$ for some
fixed $k>0$ then the $NP$-complete Problem~\ref{p:ncnm} could be
solved in polynomial time by using the following weight function for
the edges:
\begin{displaymath}
w(e)= \left\{
\begin{array}{ll}
2^{n^k} \quad \quad & \mbox{if } e \in E \mbox{ and} \\
0 & \mbox{otherwise}.
\end{array} \right.
\end{displaymath}

Thus, the problem is not approximable within $2^{n^k}$ in polynomial
time provided that $P\neq NP$.
\end{proof}

When there are only two vertices in each group the problem admits a
constant factor approximation ratio but no better:
\begin{theorem}
Problem~\ref{p:mwn} is $APX$-complete if $m = 2$. \label{t:mwn2}
\end{theorem}
\begin{proof}
This can be shown by an approximation-preserving reduction from and to
the minimum weight $2$-satisfiability problem that is known to be
$APX$-complete:
\begin{problem}[Minimum weight $2$-satisfiability~\cite{acgkmp}] \label{p:mw2sat}
Given a set $U$ of boolean variables, a set $C$ of clauses $c \in
C,|c|\leq 2$ and a weight function $w : C \to \mathbb{N}$, find the
truth value assignment $f : U \to \{ 0,1 \}$ that minimizes the sum of
the weights of unsatisfied clauses, i.e.,
\begin{displaymath}
\sum_{U \mbox{ does not satisfy } c \in C} w(c).
\end{displaymath}
\end{problem}

The reduction from Problem~\ref{p:mw2sat} to Problem~\ref{p:mwn} with
$m=2$ is very similar to the reduction in the proof of
Theorem~\ref{t:ncnm2}.  Each each boolean variable $u_i$ is
represented by a two-set $V_i=\left\{ v_i^0, v_i^1\right\}$
corresponding to truth value assignments $u_i=0$ and $u_i=1$,
respectively.  By definition of Problem~\ref{p:mwn}, the graph
$G=\left(V_1,\ldots,V_n,E\right)$ is complete, i.e., $E=\left\{
\{u,v\} : u \in V_i, v \in V_j, 1 \leq i < j \leq n \right\}$.  The
weight of the edge $e=\{ v_i^a,v_j^b\} \in E$ is zero if there is a
clause $u_i=(1-a)^2 \lor u_j=(1-b)^2$ in $C$ and $w(e)$ otherwise.

Thus, the weight of the $n$-clique $V'$ in $G$ equals to the weight of
the clauses that are not satisfied by the truth value assignment
corresponding to the $n$-clique determined by $V'$.  That is,
Problem~\ref{p:mwn} with $m=2$ is at least as difficult as
Problem~\ref{p:mw2sat}.

Problem~\ref{p:mwn} with $m=2$ can be reduced in polynomial time to
Problem~\ref{p:mw2sat} in a similar way. For each vertex set $V_i$ in
$G$ there is a boolean variable $u_i$ and the vertices $v_i^0, v_i^1
\in V_i$ correspond to the two truth value assignments of $u_i$. For
each edge $e=\{ v_i^a,v_j^b\}$ in $E$ there is a clause $u_i=(1-a)^2
\lor u_j=(1-b)^2$ with weight $w(e)$.

The weight of the clauses that the truth value assignment $f$ does not
satisfy is equal to the weight of the corresponding $n$-clique in $G$.
That is, Problem~\ref{p:mwn} with $m=2$ is at most as difficult as
Problem~\ref{p:mw2sat}.

Thus, Problem~\ref{p:mwn} with $m=2$ is $APX$-complete, as claimed.
\end{proof}

An easier variant of Problem~\ref{p:mwn} is the case where the edge
weights admit the triangle inequality within a factor $\beta$, i.e.,
for all edges $\left\{t,u \right\}$, $\left\{t,v \right\}$ and
$\left\{u,v \right\}$ in $E$ it holds
\begin{displaymath}
w(\left\{t,u\right\}) \leq \beta \left(
w(\left\{t,v\right\})+w(\left\{u,v\right\}\right)).
\end{displaymath}

A good approximation of the minimum weight $n$-clique in $G$ can be found by finding the minimum weight
$n$-star that contains one vertex from each group $V_i$. 
The method is described by Algorithm~\ref{a:mws}.

\begin{algorithm}
\caption{A constant-factor approximation algorithm for finding the minimum weight $n$-clique from a weighted graph.\label{a:mws}}
\begin{algorithmic}[1]
\Function{Minimum-Weight-Star}{$G,w$}
\State $W_{\min} \leftarrow \infty$
\For{$i=1,\ldots,n$}
 \ForAll{$v \in V_i$}
  \State $W \leftarrow 0$
  \For{$j=1,\ldots,i-1,i+1,\ldots,n$}
    \State $W \leftarrow W + \min_{u \in V_j} \left\{ w(\{u,v\}) \right\}$
  \EndFor
  \If{$W<W_{\min}$}
   \State $W_{\min} \leftarrow W$
   \State $v_{\min} \leftarrow v$
  \EndIf
 \EndFor
\EndFor
\State $V' \leftarrow \emptyset$
\For{$j=1,\ldots,n$}
 \State $V' \leftarrow V' \cup \left\{ \arg \min_{u \in V_j} \left\{ w(\{u,v_{\min}\}) \right\} \right\}$
\EndFor
\State \textbf{return} $\left(V',E'=\left\{ e \in E : e \subseteq V' \right\} \right)$
\EndFunction
\end{algorithmic}
\end{algorithm}

Algorithm~\ref{a:mws} gives constant-factor approximation guarantees
and the approximation is stable (for details on approximation
stability, see \cite{bhksu}):
\begin{theorem}
Problem~\ref{p:mwn} is polynomial-time approximable within
$2\beta\left(1-o(1)\right)$ by Algorithm~\ref{a:mws} if the edge
weights satisfy the triangle inequality within factor
$\beta$. \label{t:mwnt}
\end{theorem}
\begin{proof}
Let $G'=\left(V',E'\right)$ be the $n$-clique found from the
$n,m$-partite complete graph $G$ by Algorithm~\ref{a:mws} and let
$OPT\left(G\right)$ be the minimum weight $n$-clique in $G$.

The weight of $G'$ can be bounded above as follows.  We distribute the
weight of the solution $G'$ to its vertices:
\begin{displaymath}
w(v)=\sum_{e \in E',v \in e} w(e)/2.
\end{displaymath}
The weight of the lightest vertex in $G'$, the vertex $v_{\min}$, is
\begin{displaymath}
w(v_{\min}) = \sum_{e \in E', v_{\min} \in e} \frac{w(e)}{2} \leq
\frac{n-1}{2n\left(n-1\right)/2}OPT\left(G\right) =
\frac{1}{n}OPT\left(G\right).
\end{displaymath}
For each edge $\{u,v\} \in E'$ such that $v_{\min} \notin \{u,v\}$,
holds
\begin{displaymath}
w(\{u,v\}) \leq \beta \left[w(\left\{u,v_{\min}\right\}) +
w(\left\{v,v_{\min}\right\}) \right]
\end{displaymath}
by the assumption. Thus, the sum of the weights of the other vertices
in $V'$
\begin{eqnarray*}
\sum_{v \in V', v \neq v_{\min}} w(v) &=&
\sum_{v \in V', v \neq v_{\min}} \sum_{e \in E', v \in e} \frac{w(e)}{2} \\
& \leq&
\sum_{v \in V', v \neq v_{\min}} \sum_{e \in E', v \in e} \frac{\beta \left[w(\{u,v_{\min}\}) + w(\{v,v_{\min}\}) \right]}{2} \\
& = &
\left(n-1\right)2\beta w(v_{\min})=
\frac{2\beta\left(n-1\right)}{n}OPT\left(G\right) \\
&=&2\beta\left(1-\frac{1}{n}\right)OPT\left(G\right).
\end{eqnarray*}
Combining these two upper bounds we get
\begin{displaymath}
w(G')\leq \frac{1}{n}OPT\left(G\right) +
2\beta\left(1-\frac{1}{n}\right)OPT\left(G\right) =
2\beta\left(1-o(1)\right)OPT\left(G\right).
\end{displaymath}

Thus, Algorithm~\ref{a:mws} guarantees the approximation factor
$2\beta\left(1-o(1)\right)$ when $w$ satisfies the triangle inequality
within a factor $\beta$.
\end{proof}

This algorithm might not be applicable in orientation search as there
seems to be little hope of finding distance functions (used in
selecting the best common lines) satisfying even the relaxed triangle
inequality for the noisy projections.  However, in the case of finding
functionally analogous genes this is possible since many distance
functions between sequences are metric.  Thus, the algorithm seems to
be very promising for that task.

A very natural relaxation of the original problem is to allow small
changes to common line candidates to make the orientations consistent:
\begin{problem}[Minimum error $l$-constrained line arrangement] \label{p:mec}
Given sets $P_{ij} \subset \mathbb{R}^2, \left|P_{ij}\right|\leq l, 1
\leq i < j \leq n$, find lines $L_1,\ldots,L_n$ in $\mathbb{R}^2$ that
minimize the sum of distances $\min_{p_{ij} \in P_{ij}}
\left|p_{ij}-\hat{p}_{ij}\right|^q$ where $\hat{p}_{ij}$ is the actual
intersection point of lines $L_i$ and $L_j$ and $q >0$.
\end{problem}

Unfortunately also this variant of the problem is very difficult:
\begin{theorem}
Problem~\ref{p:mec} with $l \geq 6$ is not polynomial-time
approximable within $2^{n^k}$ for any fixed $k>0$ if $P \neq
NP$. \label{t:mec}
\end{theorem}
\begin{proof}
If Problem~\ref{p:mec} would polynomial-time approximable within
$2^{n^k}$ for some fixed $k>0$ then Problem~\ref{p:cla} could be
solved in polynomial time since the there are lines intersecting at
the allowed points if and only if the minimum error line arrangement
has error zero.
\end{proof}

\subsection{Parameterized Complexity}
Even if the problem is $NP$-hard, it might be solvable in practice if
the $NP$-hardness is caused by some properties of the inputs that do
not occur in practice.  For example, one might be interested only
vertex covers of size at most $k$.  Deciding whether there is vertex
cover of size at most $k$ in a graph of $n$ vertices can be solved in
time $O\left(n^k\right)$.  However, if $k$ is, e.g., $40$ and $n$ is
very large then this time complexity is unacceptable.  Instead, we
would like to have time complexity of form $O\left(n^c\right)$ for
some reasonably small $c$.

Formally a \emph{parameterized decision problem} is a set $D \subseteq
\Sigma^* \times \mathbb{N}$ where $\Sigma$ is a finite alphabet.  A
parameterized decision problem $D$ is \emph{fixed-parameter tractable}
if for each $\left(x,k\right) \in \Sigma^* \times \mathbb{N}$ it can
be decided whether $\left(x,k\right)$ is in $D$ in time
$f\left(k\right)|x|^{O\left(1\right)}$ where $f : \mathbb{N} \to
\mathbb{N}$ is an arbitrary function.  Parameterized complexity
classes form a hierarchy similar to the polynomial hierarchy:
\begin{displaymath}
FPT \subseteq W\left[1\right] \subseteq W\left[1\right] \subseteq \ldots \subseteq W\left[SAT\right] \subseteq W\left[P\right].
\end{displaymath}
All inclusions between the classes are believed to be proper. All
problems outside the class $FPT$ are called \emph{fixed-parameter
intractable}.

A parameterized problem $D$ reduces to a parameterized problem $D'$ if
there exist functions $f,g:\mathbb{N} \to \mathbb{N}$ and $h:D \to D'$
such that $h\left(x,k\right)$ is computable in time
$f\left(k\right)|x|^{O\left(1\right)}$ for each instance in $\Sigma
\times \mathbb{N}$, and $\left(x,k\right) \in D$ if and only if
$\left(h\left(x\right),g\left(k\right)\right) \in D'$.  Such a
reduction is called a \emph{standard parameterized $m$-reduction}.
(For further details on parameterized complexity, see~\cite{df}.)

For the orientation search problem there is a natural
parameterization: the number of projections can be bounded by a
constant. Thus, Problem~\ref{p:ncnm} can be turned into the following
parameterized problem:
\begin{problem}[$k$-clique in $k,m$-partite graph] \label{p:nckm}
Given a $k,m$-partite graph $G=\left(V_1,\ldots,V_k,E\right)$ and a
natural number $k$, decide whether there is a $k$-clique in $G$.
\end{problem}

The intuition behind this formulation of being interesting is that if
we would be able to orient a few representative projections very well
then the risk that the orientations found for the other projections
based on those well-oriented representative projections would be
incorrect could be small enough.  Thus, there would be good chances to
reconstruct an accurate density map based on the found orientations.
Unfortunately, also this formulation is fixed-parameter intractable:
\begin{theorem}
Problem~\ref{p:nckm} is $W[1]$-complete. \label{t:nckm}
\end{theorem}
\begin{proof}
Let us first show that the $k,m$-satisfiability is $W[1]$-hard:
\begin{problem}[$k,m$-satisfiability] \label{p:kmsat}
Given a set $U$ of boolean variables and a set $C,\left|C\right|=k$,
of clauses $c \in C,|c| \leq m$, decide whether there is a truth value
assignment $f : U \to \{ 0,1 \}$ that satisfies all clauses in $C$.
\end{problem}

\begin{lemma}
Problem~\ref{p:kmsat} is $W[1]$-complete. \label{t:kmsat}
\end{lemma}
\begin{proof}
The problem is shown to be $W[1]$-hard by a parameterized reduction
from the short nondeterministic Turing machine computation problem
which is known to be $W[1]$-complete.
\begin{problem}[Short nondeterministic Turing machine computation~\cite{df}] \label{p:ntc}
Given a nondeterministic Turing machine $M$, input string $x$ and a
natural number $k$, decide whether there is a computation of $M$ that
accepts the string $x$ in at most $k$ steps.
\end{problem}

It can be verified that the reduction used in Cook's Theorem (see
e.g.~\cite{hmu}) is a parameterized reduction. Thus, it can be used
also here to show that Problem~\ref{p:ntc} reduces to
Problem~\ref{p:kmsat}.

Problem~\ref{p:kmsat} can be shown to be in $W[1]$ by reduction to
Problem~\ref{p:ntc}.
\end{proof}

Problem~\ref{p:nckm} is $W[1]$-hard by a reduction from
Problem~\ref{p:kmsat} as follows.  For each clause $c_i \in C$ there
is a group $V_i$ consisting of vertices corresponding to the literals
in $c_i$.  There is an edge between $v_{i,a} \in V_i$ and $v_{j,b} \in
V_j$ if and only if $i\neq j$ and the corresponding literals can be
satisfied simultaneously.

Problem~\ref{p:nckm} can be shown to be in $W[1]$ by a reduction to
Problem~\ref{p:ntc}.
\end{proof}

\section{Conclusions \label{s:c}}
We have shown that some approaches for determining orientations for
noisy projections of identical particles are computationally very
difficult, namely $NP$-complete, inapproximable and fixed-parameter
intractable.  These results justify (to some extent) the heuristic
approaches widely used in practice.

On the bright side, we have been able to detect some polynomial-time
solvable special cases.  Also, we have described an approximation
algorithm that achieves the approximation ratio
$2\beta\left(1-o(1)\right)$ if the instance admits the triangle
inequality within a factor $\beta$.  It has promising applications in
search for functionally analogous genes.

As a future work we wish to study the usability of current state of
art in heuristic search to find reasonable orientations in
practice. This is very challenging due to the enormous size of the
search space.  Another goal is to analyze the complexity of other
approaches for determining the orientations for the projections.


%
%

\clearpage \newpage
\bibliographystyle{plain}
\bibliography{../mine}


\label{lastpage}

%
%

%
%
%
\end{document}